\documentclass[journal,twoside,shortpaper,9pt]{IEEEtran}

\usepackage{atbegshi,picture,ifpdf}
\usepackage[noadjust]{cite}
\usepackage[pdftex]{graphicx}
\usepackage[cmex10]{mathtools}
\usepackage[scaled=0.91]{helvet}
\usepackage[cmintegrals, varvw]{newtxmath}
\usepackage{bm,csquotes}
\usepackage[caption=false,font=footnotesize]{subfig}
\usepackage{url}
\usepackage[english]{babel}
\usepackage[utf8x]{inputenc}
\usepackage[T1]{fontenc}
\addto\captionsenglish{}
\usepackage{siunitx}
\newcommand*{\mj}   {\mathrm{j}}
\newcommand*{\me}   {\mathrm{e}}

\renewcommand*{\vec}{\bm}

\newcommand{\mat}[1]{\mathbf{#1}}

\newcommand{\norm}[1]{\left\lVert #1 \right\rVert}

\makeatletter
\def\ps@IEEEtitlepagestyle{%
  \def\@oddfoot{\mycopyrightnotice}%
  \def\@oddhead{\hbox{}\@IEEEheaderstyle\leftmark\hfil\thepage}\relax
  \def\@evenhead{\@IEEEheaderstyle\thepage\hfil\leftmark\hbox{}}\relax
  \def\@evenfoot{}%
}
\def\mycopyrightnotice{%
  \begin{minipage}{\textwidth}
  \centering \scriptsize
  Copyright~\copyright~2022 IEEE. Personal use of this material is permitted. Permission from IEEE must be obtained for all other uses, in any current or future media, including\\reprinting/republishing this material for advertising or promotional purposes, creating new collective works, for resale or redistribution to servers or lists, or reuse of any copyrighted component of this work in other works by sending a request to pubs-permissions@ieee.org.
  \end{minipage}
}
\makeatother

\AtBeginShipout{\AtBeginShipoutUpperLeft{%
  \put(\dimexpr\paperwidth-1cm\relax,-.55cm){\makebox[0pt][r]{
  \begin{minipage}{\textwidth}
  \centering \scriptsize 
  This is the author's version of an article that has been published in this journal. 
  Changes were made to this version by the publisher prior to publication. 
  \\[0.3ex]
  The final version of record is available at\qquad http://dx.doi.org/10.1109/TAP.2022.3145383
  \end{minipage}
  }}}%
}

\begin{document}
%
\title{{\fontsize{24}{26}\selectfont{Communication\rule{29.9pc}{0.675pt}}}\vspace*{0.2cm}\break\fontsize{16}{18}\selectfont
\vspace*{-0.1cm}%
Reliable Linearized Phase Retrieval for Near-Field\\[.75ex]Antenna Measurements with Truncated Measurement Surfaces%
}
\author{
\Large Alexander Paulus,  \IEEEmembership{\Large Graduate Student Member, IEEE}, Josef Knapp,  \IEEEmembership{\Large Graduate Student Member, IEEE},\\ Jonas Kornprobst,  \IEEEmembership{\Large Graduate Student Member, IEEE}, and Thomas F. Eibert, \IEEEmembership{\Large Senior Member, IEEE}%
\thanks{Manuscript received September 20, 2021; revised January 04, 2022; accepted January 7, 2022. Date of current
version January 8, 2021.
\emph{(Corresponding author: Alexander Paulus.)}}%
\thanks{The authors are with the Chair of High-Frequency Engineering, Department of Electrical and Computer Engineering, Technical University of Munich, 80290 Munich, Germany (e-mail: a.paulus@tum.de; hft@ei.tum.de).}%
\thanks{This work was supported in part by the German Federal Ministry
	for Economic Affairs and Energy under Grant 50RK1923.}%
\thanks{Color versions of one or more of the figures in this communication are available online at http://ieeexplore.ieee.org.}%
}
\markboth{IEEE TRANSACTIONS ON ANTENNAS AND PROPAGATION}{Paulus \emph{et al.}: Reliable Linearized Phase Retrieval for  NF Antenna Measurements}

\maketitle

\begin{abstract}
Most methods tackling the phase retrieval problem of magnitude-only antenna measurements suffer from  unrealistic sampling requirements, from unfeasible computational complexities, and, most severely, from the lacking reliability of nonlinear and nonconvex formulations.
As an alternative, we propose a partially coherent (PC) multi-probe measurement technique and an associated linear reconstruction method which mitigate all these issues.
Hence, reliable and accurate phase retrieval can be achieved in near-field far-field transformations (NFFFTs).
In particular, we resolve the issues related to open measurement surfaces (as they may emerge in drone-based measurement setups) and we highlight the importance of considering the measurement setup and the phaseless NFFFT simultaneously. 
Specifically, the influence of special multi-probe arrangements on the reconstruction quality of PC solvers is shown.
\end{abstract}


\section{Introduction}
The task of recovering the phase of a signal from measurements of its magnitudes, referred to as \textit{phase retrieval}, has received considerable attention over the past decades and the reasons for this are at least twofold.
First, phase measurements are impractical or unfeasible for a manifold of applications\,---\,ranging from the fields of optics~\cite{Gerchberg.1972,Fienup.1982}, imaging~\cite{Holloway.2016,Fogel.2016,Tian.2015}, X-ray crystallography~\cite{Pfeiffer.2006,Harrison.1993}, 
transmission (electron) microscopy~\cite{Coene.1992, Faulkner.2004b}, coherent diffraction imaging~\cite{GuizarSicairos.2008,Bacca.2019} to ptychography~\cite{Ramos.2019}. 
Hence, phase retrieval problems arise from many measurement scenarios in these fields.

Second, no \textit{truly reliable} phase reconstruction algorithm exists so far\,---\,despite the existence of a vast collection of literature in this field, e.g., related to nonconvex algorithms~\cite{Gerchberg.1972,Candes.2015,Netrapalli.2015,Pinilla.2018b,Cai.2019} 
and to convex ones~\cite{Candes.2013,Waldspurger.2015,Goldstein.2018}. 
While satisfactory results have been reported for linear measurements based on hypothetical random processes, i.e., where the measurement matrix stems from a normal distribution~\cite{Wang.2018}, existing approaches return highly sub-optimal solutions for realistic data including that of electromagnetic field measurements. 
The antenna measurement community has tried to improve the quality of the phaseless measurement data, and, thus, the suitability for phase reconstruction, by employing multiple measurement surfaces~\cite{Razavi.2010, Schmidt.2009, Yaccarino.1999,Isernia.1996,LasHeras.2020,Moretta.2019,Fuchs.2020,Varela.2021}, 
by utilizing specialized probe antennas~\cite{Costanzo.2001b,Costanzo.2005,Costanzo.2008,Paulus.2017b,Paulus.2017,Sanchez.2020}, 
by exploiting multi-frequency data~\cite{Paulus.2020,Knapp.2021}, and by considering information about spatial derivatives~\cite{Paulus.2020b}. 
While each of the attempts may yield improved results in the reported cases, there is no doubt about one fundamental flaw inherent to all existing methods. None of the approaches is \textit{guaranteed} to work when applying the same principle to a slightly different problem. 
In particular, all nonconvex solvers suffer from the problem of local minima, which might even not be recognizable as such in the presence of noise.
It follows that for all the (previously cited) algorithms in literature that they cannot judge whether a retrieved complex solution is false or close to the true solution based on the available observation data.

Despite the lack of a verifiable condition for success, the modified measurement setups and techniques empirically improve the chance of accurate phase retrieval.
Consider the following scenario for an illustration. If a measurement and phase retrieval setup, i.e., a combination of probe antennas, measurement surfaces and sampling density with a phase retrieval algorithm, is applied for the phaseless near-field (NF) characterization of two similar unknown antennas under test (AUTs), the procedure may arbitrarily fail for one set of data while it may provide accurate results for the other AUT. 
Arguably, this lack of reliability of phase retrieval algorithms prevails for a wide class of applications, including that of phaseless NF far-field transformations (NFFFTs).
So far, magnitude-only measurement techniques and reconstruction algorithms have advanced in parallel, but mostly independent from each other. 
The \textit{essential} behavior and properties of phase reconstructions for NF data have, thus, not changed\,---\,phase retrieval remains a highly nonlinear task with the described lack of reliability.

Since all approaches (so far) working with magnitude-only data are doomed to fail eventually for some input data, a simultaneous advance in  sophisticated measurement techniques and innovative phase-retrieval algorithms is, in our opinion, the preferred way to advance in order to optimally integrate all observable information into the phase reconstruction process.
We have proposed a first symbiosis between specialized measurement data and phase retrieval formalism in~\cite{Kornprobst.2021,PaulusPCEuCAP2021}. 
By assuming \textit{partial coherence} (PC) in the measurement signal, the originally nonlinear retrieval task is drastically simplified and can be solved reliably by various linearized formulations. 
In the context of phaseless NF antenna measurements, the assumption of PC can conveniently be realized at the expense of employing multi-channel receivers connected to probe antenna arrays. 
At the same time, a \textit{necessary condition} for the success of the linearized formulations exists. 
This bound on the required number of measurements is experienced to be \textit{sufficient in practice} and allows for a deterministic prediction of success or failure of the phase retrieval\,---\,an unprecedented accomplishment for phaseless NFFFTs.

In this contribution, we consider truncated NF measurements, which are for instance encountered for the in-situ characterization of AUTs by means of unmanned aerial vehicles (UAVs)~\cite{Fritzel.2016,GarciaFernandez.2017,Mauermayer.2019,Faul.2021}, 
and tackle the challenges arising for the techniques discussed in~\cite{Kornprobst.2021,PaulusPCEuCAP2021}. 
The theory of NFFFTs with incomplete phase information is discussed in Section~\ref{sec:theory}.
Subsection~\ref{ssec:phaseless_null_space} focuses on NF measurements on truncated surfaces. 
In Subsection~\ref{ssec:partially_coherent}, a linearized formulation for PC is discussed, particularly regarding possible weaknesses. 
Subsection~\ref{ssec:truncated_PC} discusses the challenges of truncated measurement surfaces.
Transformation results of synthetic NF data, based on a real-world UAV flight trajectory, are then discussed in Section~\ref{sec:results}, showcasing the advantage of the modified linearized retrieval algorithm over the unmodified variant and an existing nonlinear phase retrieval approach for PC data.
Furthermore, we demonstrate that the design of the PC measurement technique\footnote{In this work, we consider multi-probe measurement techniques. Alternatively, multi-frequency data may also be employed.} plays a crucial role for any  PC phase retrieval technique\,---\,may it be linear or nonlinear. 
This highlights the importance of our previous point that for significant advances in the field of phaseless NFFFTs, measurement techniques and algorithms have to be studied and improved simultaneously due to their inherent interdependency.

\section{Truncated Measurements with Incomplete Phase Information}
\label{sec:theory}
\subsection{Phaseless Field Transformation}
\label{ssec:phaseless_null_space}
A phaseless NFFFT can be formulated as finding the unknown coefficients $\vec z\in\mathbb{C}^{n}$ of a known AUT representation as solution of
\begin{align}
	\left|\mat A \vec z\right|^2 &= \left|\vec b\right|^2,\label{eq:phaseless_main}
\end{align}
where the measurement matrix $\mat A\in\mathbb{C}^{m\,\times\,n}$ represents the linear relationship between the AUT coefficients and the complex-valued measurement vector $\vec b\in\mathbb{C}^{m}$ acquired by the probe antennas. 
The magnitude operator $|\cdot|$ and the exponent $(\cdot)^2$ act element-wise on any vector.
Based on the uniqueness and surface equivalence theorem~\cite{Jin.2015}, the \textit{complete} far field (FF) of the AUT can uniquely be determined from measurements of the tangential fields acquired on a \textit{closed} surface surrounding the AUT. 
The functional principle of an equivalent-source-based NFFFT contains two steps.
First, an equivalent representation,\footnote{The specific choice of the equivalent source representation does not matter much as long as all degrees of freedom of the AUT are captured~\cite{Kornprobst.2021b}.} i.e., vector spherical wave functions or current densities, of the AUT is determined from the probe signals acquired on the measurement surface(s).
Second, the FF of these equivalent sources is evaluated, which equals the FF of the AUT. 
The absence of phase information does only alter the first part of the transformation. With full phase information, a linear system of equations has to be solved. When incomplete or no phase information is available, the nonlinear system of equations in~\eqref{eq:phaseless_main} is encountered and a phase retrieval procedure is required. 

Whenever NF measurements on \textit{incomplete} measurement surfaces are acquired, e.g., on single planar or cylindrical surfaces, the reconstructed FF in the direction of the regions lacking measurement data is in general not valid. As the fields radiated in these directions do not have to be represented by the equivalent sources, the latter may be modified, i.e., reduced or truncated, leading to a smaller number of unknowns. However, it is not strictly necessary to modify the equivalent sources according to the measurement geometry and often not even possible without complicating the overall approach, e.g., in case of spherical vector wave functions~\cite{Hansen.2008}. 
In any case, it can happen that the measurement samples in the truncated measurement region are no longer sufficient to \textit{uniquely} determine, i.e., restrict, the equivalent sources and that so-called \textit{non-radiating sources} $\vec z_{\mathrm{nr}}$ may occur. From a mathematical point of view, these sources reside in the null space of the operator $\mat A$, i.e., $\mat A \vec z_{\mathrm{nr}} = \vec 0$, and are, thus, not observable by the probe antennas.
The term \enquote{non-radiating} only refers to fields at the measurement locations. While the sources do not generate any fields in the (truncated) measurement region, it is likely that they generate significant fields outside the actual measurement region. 
This artificial null space associated with truncated NF measurement setups can be avoided in at least three ways. First, a change of basis, e.g., via a truncated singular value decomposition (SVD), may be applied, allowing to represent the AUT in terms of radiating and non-radiating currents with respect to the available measurement samples, where only the radiating ones are employed in the solution process. 
Second, a regularized solver can be employed, which keeps the $\ell_2$-norm of the source coefficients small such that unobservable currents are suppressed. 
Third, artificial measurement entries of zero value can be added within the regions which have not been covered in the measurement process. By adding the corresponding rows in the forward operator, the dimension of the potential null space is reduced and a non-trivial null space in the forward operator (with influence on the radiated fields) is effectively avoided.
The spatial separation between the original samples and the artificial zero-field locations has to be chosen carefully. While a narrow gap ensures elimination of most parts of the null space, it represents a physically impossible field distribution and may deteriorate the accuracy of the retrieved FF in the valid radiation directions.

\subsection{Partially Coherent Transformation}
\label{ssec:partially_coherent}
In line with the ideas presented in~\cite{Kornprobst.2021,PaulusPCEuCAP2021}, the availability of incomplete phase information in the measurement vector $\vec b$ is assumed and a linearized phase retrieval algorithm is derived. 
We assume a simplistic scenario of PC, which allows a convenient introduction of the underlying equations. 

Suppose that we have measured an AUT twice,  once with a single probe antenna and a single-channel receiver yielding $\vec b_1\in\mathbb{C}^{m_1}$, and a second time with a two-element probe antenna array connected to a two-channel receiver resulting in $\vec b_2\in\mathbb{C}^{m_{2}}$ and $\vec b_3\in\mathbb{C}^{m_{2}}$, where the two-element probe array is placed at $m_2$ sample locations. 
We further require that the two-channel receiver, measuring $\vec b_2$ and $\vec b_3$, is able to observe the phase difference between its two channels\,---\,which is possible with scalar receivers via linear combinations (LCs), e.g.,~\cite{Costanzo.2001b,Costanzo.2005,PaulusPCEuCAP2021}, or a vectorial device with coherent channels, e.g., a standard multi-channel vector network analyzer or a suitable software defined radio~\cite{Mauermayer.2019}. 
It follows that the phase differences between the $k$th entries in $\vec b_2$ and $\vec b_3$ for $k\in\left\{1,...,m_2\right\}$ is, thus, known. 

Without loss of generality, we consider the particular values of $m_1 = 1$, $m_2 = 2$ for the measurement vectors
\begin{align}
	\vec b_1 = \left|b_{11}\right|\me^{\,\mj \varphi_{11}},\hspace*{0.07cm}
	\vec b_2  = \begin{bmatrix}
		\left|b_{21}\right|\me^{\,\mj \varphi_{21}} \\ \left|b_{22}\right|\me^{\,\mj\varphi_{22}}
	\end{bmatrix},
	\hspace*{0.07cm}
	\vec b_3 = \begin{bmatrix}
		\left|b_{31}\right|\me^{\,\mj \varphi_{31}} \\ \left|b_{32}\right|\me^{\,\mj\varphi_{32}}
	\end{bmatrix},\label{eq:example}
\end{align}
which include observed magnitudes and unknown phase terms $\me^{\,\mj \varphi_{ji}}$. 
However, the complex exponentials of phase differences $\me^{\,\mj \left(\varphi_{31} - \varphi_{21}\right) }$ and $\me^{\,\mj \left(\varphi_{32} - \varphi_{22}\right) }$ are observable by the assumed two-channel receiver.
This fact is exploited when we formulate the inverse-source problem
\begin{align}
\mat A&\mat A^{\dagger} \vec b	=\mat A\vec z = \vec b = \begin{bmatrix}
		\vec b_1^{\text{T}} & \vec b_2^{\text{T}} & \vec b_3^{\text{T}} 
	\end{bmatrix}^{\text{T}} \nonumber\\&= \begin{bmatrix}
		\left|b_{11}\right|   &    0                   & 0                     \\
		0                     &  \left|b_{21}\right|   & 0                     \\
		0                     &  0                     & \left|b_{22}\right|   \\
		0                     &  \left|b_{31}\right|\me^{\,\mj \left(\varphi_{31} - \varphi_{21}\right) }   & 0                     \\
		0                     &  0                     & \left|b_{32}\right| \me^{\,\mj \left(\varphi_{32} - \varphi_{22}\right) }  \\
	\end{bmatrix}\begin{bmatrix}
		\me^{\,\mj \varphi_{11}} \\ \me^{\,\mj \varphi_{21}} \\\me^{\,\mj \varphi_{22}}
	\end{bmatrix}\nonumber \\
	&= \text{diag}\left(\left|\vec b\right|\right) 
	\begin{bmatrix}
		1   &    0                   & 0                     \\
		0                     &  1   & 0                     \\
		0                     &  0                     & 1   \\
		0                     &  \me^{\,\mj \left(\varphi_{31} - \varphi_{21}\right) }   & 0                     \\
		0                     &  0                     &  \me^{\,\mj \left(\varphi_{32} - \varphi_{22}\right) }  \\
	\end{bmatrix} \vec \psi\nonumber \\
	&= \mat B \mat C \vec \psi,\label{eq:intermediate_PC}
\end{align}
where the remaining phase unknowns of only $\vec b_1$ and $\vec b_2$ are stacked in the vector $\vec \psi\in\mathbb{C}^{q}$ (with $\left|\vec \psi \right| = \vec 1$, $\vec 1$ being a suitable all-ones vector), and $\mat A^{\dagger}$ is the pseudo inverse of $\mat A$. 
The operator $\mathrm{diag}(|\vec b|) =\mat B\in\mathbb{R}^{m\times m}$ constructs a diagonal matrix from the observed magnitudes and the known phase differences are placed inside the phase-difference matrix $\mat C\in\mathbb{C}^ {m\times q}$.
Comparing the farmost left-hand and right-hand sides of~\eqref{eq:intermediate_PC} reveals that we have been able to eliminate the source coefficients $\vec z$.
This becomes more obvious by rewriting~\eqref{eq:intermediate_PC} as
\setlength{\belowdisplayskip}{6pt}
\setlength{\abovedisplayskip}{6pt}
\begin{align}
	(\mat I - \mat A\mat A^{\dagger})\mat B \mat C \vec \psi &= 0 \label{eq:PC_linear_psi_main}\\
	\text{s.t.}\,\, [\vec \psi]_s = 1. \nonumber
\end{align}
In the side constraint of~\eqref{eq:PC_linear_psi_main}, we have already relaxed the nonconvex condition on all of the magnitudes $|\vec \psi | = \vec 1$ of the remaining phase unknowns to only affect the single $s$th entry $[\vec \psi]_s$, which represents a convex restriction. 
In essence,~\eqref{eq:PC_linear_psi_main} is an alternative formulation to that in (10) of~\cite{PaulusPCEuCAP2021}. Here, a reduced number of unknowns at the expense of a larger computational effort due to the occurrence of the pseudo inverse matrix $\mat A^{\dagger}$ is achieved compared to the formulation in~\cite{PaulusPCEuCAP2021}. Note that both formulations have, in a different notation, already been mentioned in~\cite{Kornprobst.2021}. %

The idea of~\eqref{eq:PC_linear_psi_main} is to determine the remaining unknown phases in the vector $\vec \psi$ such that no \textit{physically incorrect} portions are generated, which might contradict the source model, when combined with the known phase differences in the matrix $\mat C$ and the measured magnitudes in $\mat B$. In other words, whenever the combination of phases with magnitudes can be generated by the underlying model, i.e., by the forward operator $\mat A$, the term on the left-hand side in the equation in~\eqref{eq:PC_linear_psi_main} will be zero. The projector $\mat A \mat A^{\dagger}$ removes portions which do not agree with the equivalent model, while the identity matrix $\mat I$ leaves any vector unchanged\,---\,the difference $\mat I \tilde{\vec b} - \mat A\mat A^{\dagger}\tilde{\vec b}$ will, thus, reveal inappropriate portions in $\tilde{\vec b}$.

\subsection{Considerations for Truncated Measurement Surfaces\label{ssec:truncated_PC}}

We have shown in~\cite{Kornprobst.2021} that two formulations related to~\eqref{eq:PC_linear_psi_main} have a unique, non-trivial homogeneous solution, i.e., the dimension of the null space of the involved operator (including the projection matrix) is one.
A necessary but not sufficient condition for a unique solution was derived as~\cite{Kornprobst.2021}
\begin{align}
	m-\mathrm{rank}\left(\mat{A}\right) \ge  \mathrm{rank}\left(\mat B \mat C\right)-1 \label{eq:PC_necc_bound_final2}
\end{align}
dependent on the rank 
of the forward matrix and the product $\mat B \mat C$. 
For any realization of the phase retrieval concept, we must ensure that the dimensionality of the null space remains one. 
This may be checked in general for the operator of~\eqref{eq:PC_linear_psi_main} $(\mat I - \mat A\mat A^{\dagger})\mat B \mat C$ (which contains the measured magnitudes and phase differences) only after the measurements have been taken.

With reference to the discussion of truncated measurement surfaces in Subsection~\ref{ssec:phaseless_null_space}, we already know beforehand that such a problem may occur.
Imagine that the operator $\mat A$ exhibits a non-trivial null space by construction, e.g., caused by a truncated NF measurement setup. 
Hence, we know that the operator $\mat A$ allows the existence of non-trivial sources $\vec z _\mathrm{nr}$ which are not observable at the measurement surface but do radiate in other (NF and FF) regions.
When solving~\eqref{eq:PC_linear_psi_main}, intermediate quantities in the source domain are computed, e.g., 
\begin{equation}
\vec z^\prime = \mat A^{\dagger}\mat B \mat C \vec \psi = \vec z + \vec z_{\mathrm{nr}}.
\end{equation}
Note that the product of $\mat B \mat C$ is ensured to have full rank by construction, i.e., $\mathrm{rank}(\mat B \mat C) = q$, if $\vec b$ does not contain zero entries.
Thus, only the kernel of $\mat A$ is of interest. 
Having $\vec z_\mathrm{nr}$ with an arbitrary scaling is a valid homogeneous solution of~\eqref{eq:intermediate_PC}. 
This is a serious issue since the whole linearization of the phase retrieval problem is based on retrieving a unique non-trivial null-space vector. If not taken care of properly, this may render the linearized phase retrieval strategy useless.
When the source quantities are affected by non-radiating sources, the energy in $\vec z_{\mathrm{nr}}$ may grow extremely large in comparison to the desired solution. 
This depends to some extent on the employed solver. 
However, with an increasing homogeneous part in the source vector $\vec z'$, the NF radiation in the truncation region grows larger. Eventually, the FF is dominated by contributions in the truncation regions and, as a consequence, numerical cancellation errors and leakage of the huge radiation in the truncation region into the valid measurement region may adversely affect the solution of~\eqref{eq:PC_linear_psi_main}.

\begin{figure}[t]
	\centering%
	\includegraphics{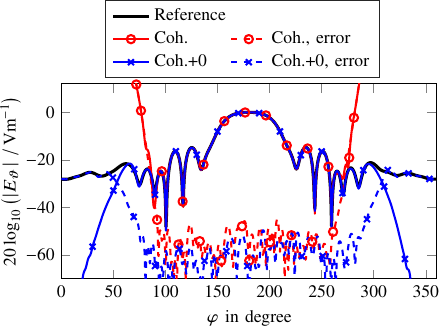}
	\caption{Impact of non-radiating sources on the reconstructed FF from truncated NF measurements for a direct solver with full coherence (coh.) and without regularization. Explicitly enforcing zero fields in the unsampled NF regions limits the FF outside the valid region (\enquote{+0}). }
	\label{fig:null_space_FF}
\end{figure}

Figure~\ref{fig:null_space_FF} illustrates the impact of an artificial null space in the forward operator on the FF obtainable with a fully coherent transformation. A truncated cylindrical NF measurement setup is employed (identical to the one discussed in Section~\ref{sec:results}) and the transformation is performed via a direct solver and without regularization of the equivalent sources. In line with the discussion above, the fields outside the sampled region are not bound, such that non-radiating currents are excited during the solution process. These currents generate arbitrary, potentially large fields outside the valid region, as seen in Fig.\,\ref{fig:null_space_FF}. By artificially enforcing the field to be zero in the unsampled NF areas, the non-trivial null space is suppressed, the trivial solution is enforced in the truncation region, and non-radiating currents are avoided. The procedure leads to the curves labeled with the suffix \enquote{+0} in Fig.\,\ref{fig:null_space_FF} and is explained in more detail in the results section.
As observed, the unbounded radiation negatively affects the FF in the ranges $50^\circ<\varphi<100^\circ$ and $250^\circ<\varphi<300^\circ$ for the considered $\vartheta=90^\circ$ cut, and, overall, the accuracy in the valid measurement region (approximately $70^\circ < \varphi < 290^\circ$, see Fig.~\ref{fig:ALUTSA_antenna}) is increased for the case \enquote{+0}.

\section{Transformation of Drone-Based Data with Partial Coherence}
\label{sec:results}
\begin{figure}[t]
	\centering%
	\includegraphics{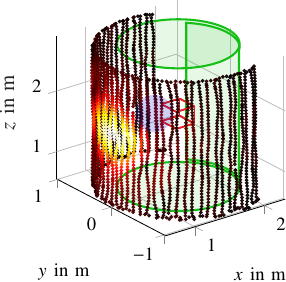}
\caption{A synthetic NF antenna measurement setup combining the ALUTSA as an AUT with a real-world drone trajectory for the definition of the measurement locations and probe orientations. The ALUTSA currents for NF generation are located on the red box, whereas for the field transformations equivalent surface current densities on an enclosing sphere, indicated in blue, are assumed. The green areas mark regions which have not been sampled and which may require special treatment within the phaseless transformation process.}
	\label{fig:ALUTSA_antenna}
\end{figure}

The effect of truncated NF measurements and the inherent null space in the forward operator on the performance of the linearized phase retrieval solver in~\eqref{eq:PC_linear_psi_main} for PC is investigated for synthetic data on a real-world UAV flight trajectory. 
A simulation model of the aluminum tapered-slot antenna (ALUTSA), described in~\cite{Azhar.2021}, was employed to generate NF data, and the sampling locations were placed on a UAV flight trajectory acquired with the multicopter and positioning system described in~\cite{Mauermayer.2019}. 
In this way, NF data for arbitrary probe antennas at realistically and irregularly distributed acquisition points could be generated. 
Note that the probe orientation angles were taken from the drone data set (6D position data).
The cylindrical flight path and the resulting field values of the co-polarization of the ALUTSA model are drawn in Fig.\,\ref{fig:ALUTSA_antenna}.
\setlength{\floatsep}{5pt plus 0pt minus 2pt}
\addtolength{\textfloatsep}{-0.14in}
The radiation of the ALUTSA is here modeled by equivalent sources placed on an enclosing box, drawn in red. In order to avoid inverse crime in the field transformations, $n = \num{3e3}$ Hertzian dipoles placed on the surface of the blue sphere around the original model are utilized as the equivalent source representation for the transformation. Field regions which have not been sampled in the course of the truncated cylindrical measurement have been marked in green color in Fig.\,\ref{fig:ALUTSA_antenna}. 

\begin{figure}[t]
	\centering%
\hfill
	\subfloat[\label{fig:ALUTSA_probe_jonas}]{%
	\includegraphics{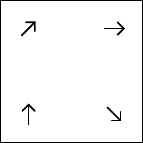}
   }
\hfill
\subfloat[\label{fig:ALUTSA_probe_bad}]{%
	\includegraphics{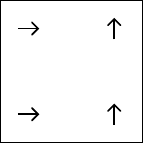}%
	}
\hspace*{\fill}
\caption{Probe antenna arrays employed in combination with the virtual measurement setup of Fig.\,\ref{fig:ALUTSA_antenna}. The two-dimensional arrays consist of four Hertzian dipoles, which sample the tangential electric field within the plane of the probe. The arrows indicate the acquisition of the vertical, horizontal or an equally weighted sum of both field components. The horizontal and vertical spacing between the elements is approximately one wavelength.}
\label{fig:probe_arrays}
\end{figure}

For the given drone locations and orientations, the two probe antenna arrays illustrated in Fig.\,\ref{fig:probe_arrays} haven been employed for data generation. 
They consist of four elements each, here Hertzian dipoles, which sample the vertical and horizontal component of the electric field, or an equally weighted sum of both. In the following, it is assumed that the four elements are connected to a four-channel receiver, resulting in partial coherence among the elements at each sample location. When the probe array is moved to the next point on the drone trajectory, a new quadruple of signals is recorded which features no coherence to the previous one.
Note that chain-linking of the phase data (which has been proposed in~\cite{Costanzo.2001b}) is not feasible due to several reasons. The subsequent measurements are not aligned in location and orientation, and even if that would be the case, the measurement probes themselves have a varying polarization angle.

\begin{figure}[t]
	\centering%
	\subfloat[\label{fig:ALUTSA_probe_jonas_cNF}]{%
	\includegraphics{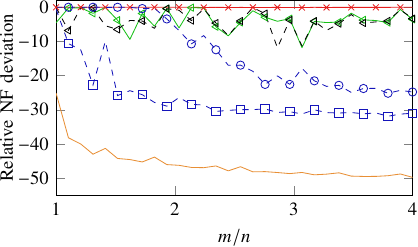}
		}
	\\[1ex]%
		\subfloat[\label{fig:ALUTSA_probe_bad_cNF}]{%
	\includegraphics{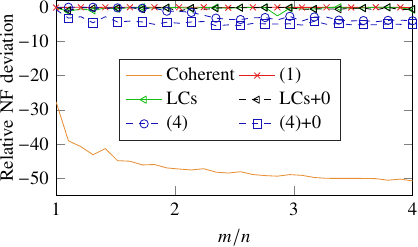}
	}%
		
	\caption{The obtained NF deviation when varying the number of measurements for $n_{\text{SNR,dB}} = \SI{60}{\dB}$ and when utilizing the arrays in Figs.\,\ref{fig:probe_arrays}(a) and~\ref{fig:probe_arrays}(b), respectively.}
\label{fig:ALUTSA_proberesults}
\end{figure}

The achievable relative NF deviation
\begin{align}
	\epsilon_{c,\mathrm{dB}}\left(\mat A \vec z, \vec b\right) &=20\,\text{log}_{10} \left( \dfrac{\norm{\mat{A}\vec{z}-\vec{b}}_2}{\norm{\vec{b}}_2}\right)
\end{align}
for a varying number of measurements $m$, which were randomly picked from the available locations on the drone trajectory, is shown in Fig.\,\ref{fig:ALUTSA_proberesults}(a) for the probe in Fig.\,\ref{fig:probe_arrays}(a) and in Fig.\,\ref{fig:ALUTSA_proberesults}(b) for the array in Fig.\,\ref{fig:probe_arrays}(b). For both figures, noise according to a signal-to-noise ratio (SNR) of $\SI{60}{\dB}$ defined via
\begin{align}
	n_{\text{SNR}} &= \left[\dfrac{\text{max}\left(\left|\vec b\right|\right)}{\text{std}\left(\vec n\right)}\right]^2\,\,\text{and}\,\,\,n_{\text{SNR,dB}} = 10\,\text{log}_{10}\left(n_{\text{SNR}} \right) \label{eq:SNR}
\end{align}
was added to the measurement vectors.

Six field transformation algorithms were employed, including as a reference a transformation utilizing full coherence. A nonconvex and nonlinear phase retrieval method~\cite{Paulus.2017b} operating on~\eqref{eq:phaseless_main} was added as a characteristic representative of common nonlinear reconstruction methods for the case of complete incoherence. The same formulation was also employed for the incorporation of the phase differences in the form of the magnitudes of LCs, analogously described in (4) of~\cite{PaulusPCEuCAP2021} and results are shown for comparison. The latter as well as the linearized formulation from~\eqref{eq:PC_linear_psi_main} have been applied in two variants, one solely operating with the provided measurement vectors including PC, and a second one with additional artificial measurements of zero magnitude within the unsampled regions (labeled with \enquote{+0}). 
The initial guess for all nonconvex solvers was generated according to the optimal spectral method~\cite{Mondelli.2019,Luo.2019b} via the implementation provided in the PhasePack library~\cite{Chandra.2019,Chandra.2017}.
By enforcing a zero field in the regions marked in green in Fig.\,\ref{fig:ALUTSA_antenna}, the null space of the forward operator is dampened and the occurrence of non-radiating currents is suppressed. In Fig.\,\ref{fig:ALUTSA_proberesults}(a) this leads to a reduced NF deviation for the linearized solver, whereas for the nonconvex solver employing LCs no significant difference is visible. Overall, Fig.\,\ref{fig:ALUTSA_proberesults}(a) indicates the superior performance of the linear formulations compared to the state-of-the-art nonconvex approaches throughout the range of $m/n$ values for an SNR of $\SI{60}{\dB}$.  Still, the accuracy of a transformation with full phase information is not achieved by any of the PC-solvers.

Despite the similarity of the probe arrays at first glance, the results in Fig.\,\ref{fig:ALUTSA_proberesults}(b) obtained for the probe in Fig.\,\ref{fig:probe_arrays}(b) are significantly worse for the solvers operating on incomplete phase information. Looking closer at the probe arrays, it becomes evident that the second probe array does not provide phase difference information along the horizontal direction for each polarization\,---\,only in vertical direction. 
In contrast, the array in Fig.\,\ref{fig:probe_arrays}(a) provides connections in horizontal and vertical direction via its $\SI{45}{\degree}$-rotated array elements. 
This seemingly minor difference may have considerable impact on the performance of a variety of phase retrieval solvers, not limited to the discussed linearized formulation.
We conclude from this observation that the advance on either the algorithmic or the measurement part alone is not sufficient for improved phase retrieval techniques for NFFFTs. Both aspects have to be considered at the same time, and understanding their interaction is vital in order to attain solutions for truly reliable phase retrieval.

\begin{figure}[t]
	\centering%
	\subfloat[]{%
	\includegraphics{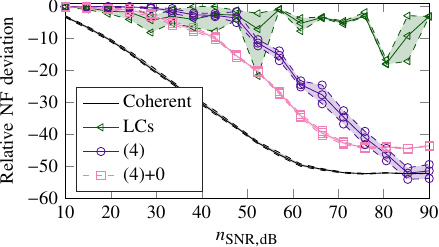}
		\label{fig:ALUTSA_probe_jonas_cNF_SNR_sweep}%
	}%
	\\[1ex]%
		\subfloat[]{%
	\includegraphics{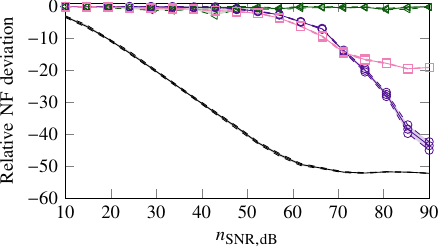}
	\label{fig:ALUTSA_probe_bad_cNF_SNR_sweep}%
}%
	\caption{Achievable NF deviation for varying SNR values. For each SNR value $20$ repetitions with random noise realizations and with $m/n \approx 3$ were conducted. The results in (a) and (b) are based on the probe antenna array of Figs.\,\ref{fig:probe_arrays}(a) and~\ref{fig:probe_arrays}(b).}
	\label{fig:ALUTSA_proberesults_SNR_sweep}
\end{figure}

A clearer picture of the impact of the choice of probe and the achievable improvements when adding the zero-measurements can be gained from Fig.\,\ref{fig:ALUTSA_proberesults_SNR_sweep}. 
For a fixed sampling ratio of $m/n\approx 3$ 
, $20$ random noise realizations have been considered for a varying SNR. As the additional measurements have not been seen to improve the comparison solver, the approach \enquote{LCs+0} was excluded from this SNR study. 
Also, the pure phaseless solver~\eqref{eq:phaseless_main} was excluded due to its complete lack of successful phase retrieval for the considered scenario.
Depicted is the minimum, maximum and arithmetic average of the relative NF error which was obtained for each SNR value, drawn with dashed lines and a solid curve, respectively. 
In Fig.\,\ref{fig:ALUTSA_proberesults_SNR_sweep}, the suppression of the artificial null space improves the noise robustness of the linearized solver, until at larger SNR values the maximal accuracy is limited. Enforcing a zero field on the measurement-free regions
ultimately yields an unphysical condition, which cannot be fulfilled exactly by the equivalent sources. 
However, this limitation does only become relevant once the noise-induced errors decrease below this particular level of precision. 
The overall insufficient performance of the probe array of Fig.\,\ref{fig:probe_arrays}(b) does not improve notably when introducing the artificial null samples. The lack of horizontally directed phase differences for either polarization makes the approach far more sensitive to noise and other issues, e.g., the additional null space, can be considered secondary. 
This casts doubts at the general applicability of multi-probe solutions for phaseless NF measurements reported in literature, whose array elements are mostly placed in one dimension only~\cite{Costanzo.2005,Paulus.2017b,Paulus.2017,Sanchez.2020}.

\section{Conclusion}
We have presented a linear phase retrieval algorithm which can reliably reconstruct the absolute phase from partially coherent measurement data. 
An additionally increased null-space dimension associated with truncated measurement surfaces may cause significant accuracy problems for the proposed linear method, which relies on retrieving a unique non-trivial null-space vector of a linear system of equations constructed for PC. 
The problems can be cured to a certain extent by introducing artificial null samples in the measurement-free regions. This work extends the applicability of the reliable linear phase reconstruction method presented in previous works to open measurement surfaces, which are found in many practically relevant measurement scenarios.

Our investigations have shown that phase retrieval algorithms with PC should always be studied together with the measurement setup, since the influence of the multi-probe arrangement (or, more generally, the PC measurement technique) is rather significant for the outcome. In general, the absence of phase information in any form was seen to render field transformation algorithms more sensitive with respect to noise. In direct consequence, phaseless measurements need to be carried out with higher precision, e.g., larger SNR and lower positioning uncertainty, than their coherent counterparts in order to obtain results of similar accuracy. 
However, also in the more challenging case discussed in this paper, the reliability can only be ensured by the proposed linearized PC phase retrieval algorithm.
Eventually, this increased reliability justifies its rather large computational effort (consisting of a nested pseudo-inversion inside a linear system of equations). 
For PC NFFFTs, the probe design and further algorithmic improvements for low-SNR measurements remain topics for future research, though.

\bibliographystyle{IEEEtran}
\bibliography{IEEEabrv,selected_phaseless}

\end{document}